\documentstyle[graphics,twocolumn]{mn}
\newcommand{\mincir}{\raise
-2.truept\hbox{\rlap{\hbox{$\sim$}}\raise5.truept 
\hbox{$<$}\ }}
\newcommand{\magcir}{\raise
-2.truept\hbox{\rlap{\hbox{$\sim$}}\raise5.truept
\hbox{$>$}\ }}
\newcommand{\minmag}{\raise-2.truept\hbox{\rlap{\hbox{$<$}}\raise
6.truept\hbox
{$>$}\ }}

\newcommand{\lya}{Lyman-$\alpha$~}

\newcommand{\be}{\begin{equation}}
\newcommand{\ee}{\end{equation}}
\newcommand{\ba}{\begin{eqnarray}}
\newcommand{\ea}{\end{eqnarray}}
\newcommand{\brr}{\begin{array}}
 
\newcommand{\err}{\end{array}}
\newcommand{\bc}{\begin{center}}
\newcommand{\ec}{\end{center}}

\newcommand{\cm}{\,{\rm cm}}

\newif\ifAMStwofonts
\AMStwofontstrue

\DeclareMathAlphabet{\mathsc}{OT1}{cmr}{m}{sc}
\def\testbx{bx}%
\DeclareRobustCommand{\ion}[2]{%
\relax\ifmmode
\ifx\testbx\f@series
{\mathbf{#1\,\mathsc{#2}}}\else
{\mathrm{#1\,\mathsc{#2}}}\fi
\else\textup{#1\,{\mdseries\textsc{#2}}}%
\fi}

\title[The effect of (strong) discrete absorption systems on the 
\lya forest flux power spectrum]{The effect of (strong) discrete 
absorption systems on the \lya forest flux power spectrum
}
\author[M. Viel, M.G. Haehnelt,  R.F. Carswell, T.-S. Kim] 
{M. Viel, M.G. Haehnelt, R.F. Carswell, T.-S. Kim 
\\ 
Institute of Astronomy, Madingley Road, Cambridge CB3 0HA\\
\\}

\begin{document}

\maketitle

\begin{abstract}
We demonstrate that the  \lya forest flux power spectrum  of ``randomised'' 
QSO absorption spectra is comparable in shape  and amplitude to the 
flux power spectrum of the original  observed spectra. In the 
randomised spectra a  random shift  in  wavelength has been added to 
the observed absorption lines as identified and fitted with VPFIT. 
At  $0.03$ s/km  $<k<0.1$ s/km  the ``3D'' power spectrum  of the 
randomised flux  agrees with that of  observed spectra  within 
the errors.  At larger scales  it is
still  $\ga 50 \% $ of that of  the  observed spectra.  At smaller 
scales the flux power spectrum is dominated by metal lines.  Lines
of  increasing column density contribute to the ``3D'' flux power
spectrum  at increasingly larger scales. Lines with $13<\log 
(N_{\rm HI}/\cm^{-2})<15$   dominate at the peak of the ``3D'' power 
spectrum  while  strong absorbers with 
$\log (N_{\rm HI}/\cm^{-2}) >15$ 
dominate at large scales,  $k< 0.03$ s/km. We further show that a fraction
of up to 20\%  of the mean flux decrement  is
contributed by strong absorbers.
Analysis  of the flux power spectrum which use 
numerical simulations  with too few strong absorption systems 
calibrated with  the observed mean flux 
are likely to underestimate the 
inferred {\it rms} fluctuation amplitude  and the slope of the initial
DM matter power  spectrum.  
In a combined analysis with other data  which constrains the DM 
power spectrum on larger scales this   can result  
in a  spurious detection of a running spectral index.
\end{abstract}


\section{Introduction}
          
In the mid  1990s a paradigm shift occurred in the interpretation 
of the \lya forest. Instead of being caused by small (kpc size) 
``\lya  clouds'' it is now widely believed  that most of the
absorption  arises from smooth fluctuations in the density of 
a photoionized warm intergalactic medium (see Rauch 1998 
and Weinberg 1999 for reviews).  Traditionally 
absorption spectra had been decomposed into Voigt profiles 
which have then been identified with individual discrete absorption
systems. For these absorbers column density and Doppler parameter 
distribution  and correlation function were determined
(Rauch 1998). With the new   paradigm  
the emphasis of  the analysis has shifted to  
statistical measures more suitable for absorption arising from 
a continuous density field, most notably the flux decrement
distribution  and the flux power spectrum.  While the  clustering
signal in the correlation function  of discrete absorbers was very weak
for all but  the  strongest  absorption systems (Cristiani et al. 1995) 
the flux power spectrum indicated {\it rms} fluctuations of 10-30\% on scales 
of a few Mpc  decreasing  as a power-law towards larger scales.  
This was generally interpreted as a detection of the clustering of the 
matter distribution. Due to non-linear and saturation effects the
relation to the dark matter(DM) power spectrum is not straightforward
even on large scales. Numerical simulations 
are therefore used to constrain  amplitude and slope of the matter 
power spectrum.  The constraints are broadly consistent with the so called 
concordance model of structure formation (Croft et al. 1999, 
McDonald  et al. 2000, Croft et al. 2002, Kim et al. 2003 (K03), Viel
et al. 2003).
Recently a  combined analysis of \lya forest and  WMAP data has been 
used to argue that a   spectral index  of the initial matter
fluctuation spectrum $<1$ and/or a running running spectral index 
is indicated by the data (Spergel et al. 2003, Verde et
al. 2003).  The errors are still large (Seljak, McDonald \& Makarov
2003)  and the  hope is that tighter constraints can be obtained  
from the SDSS sample of  QSO absorption spectra 
(Mandelbaum et al. 2003) . However,  
the reason why the clustering signal in the correlation
function of weak absorption systems is apparently 
so much weaker than that in the flux power spectrum has not been 
investigated in detail.    Press, Rybicky \& Schneider (1993) 
have demonstrated 
that the  {\it rms} fluctuation of the flux distribution on scales of
25 \AA\  
due to randomly 
distributed  absorption lines is  comparable to the observed
fluctuations.  Zuo \& Bond (1994) have shown that the flux correlation
function of low-resolution absorption spectra can be reproduced 
by a superposition of randomly distributed absorption lines.  
It is thus  astonishing that 
the flux power spectrum of a  random distribution of
absorption systems  has not yet been investigated.

Here we have studied the flux power spectrum of a 
sample of  8 randomised high-resolution 
absorption spectra taken with the  VLT-UVES  spectrograph for  which 
a complete decomposition into Voigt 
profiles due to hydrogen and metal lines absorption lines 
had been performed previously (Kim et al. 2001, Kim 
et al. 2002).  In section 2 we describe the data. Section 3 compares 
the flux power spectrum of randomised and observed spectra. Section 4 
investigates  the effect of strong absorption systems. In Section 5 we 
discuss consequences  for  the inferred DM power spectrum and in Section
6 we give our conclusions.

\section{The Data}

The sample consists of 8 spectra taken with the Ultra-Violet Echelle
Spectrograph (UVES) on VLT. The 8 spectra were taken from the ESO
archive.   The median redshift of the sample is $<z>=2.43$
In Table \ref{tab1} we list the \lya redshift range, wavelength range
and signal-to-noise ratio of the sample. 
The data reduction is described in Kim et al. (2001, 2002, 2003). 
Line lists for hydrogen and metal absorption have been compiled with the 
fitting routine  VPFIT (Carswell et al.: 
http://www.ast.cam.ac.uk/~rfc/vpfit.html). For each of the 8 spectra 
we have produced randomised spectra  where we have randomly shifted 
the hydrogen lines in 
wavelength.  This procedure results in spectra 
with a column density distribution which is very similar to  
that of the  observed spectra.  Note, however, that there will be some 
incompleteness  at the low column density end due to lines which have 
not been identified by the fitting procedure due to blending.  We have 
also produced   randomised spectra where we restricted the input list
of  hydrogen lines to certain column density ranges.

\section{The flux power spectrum of randomised spectra} 

\subsection{Calculating the flux power spectrum} 

\begin{table}
\caption{QSO absorption spectra}
\label{tab1}
\begin{tabular}{lccc}
\hline
\noalign{\smallskip}
QSO & $z_{\mathrm{Ly\alpha}}$ & $\lambda_{\mathrm{Ly\alpha}}(\rm{\AA})$
& S/N\\
\noalign{\smallskip}
\hline
\noalign{\smallskip}
Q0055--269   & 2.93--3.61 & 4778--5603 &30-75\\
Q0302--003   & 2.95--3.24 & 4807--5156 &55-75\\
HE2347--4342 & 2.29--2.84 & 4002--4669 &40-60\\
Q1101--264   & 1.65--2.11 & 3224--3780 &30-70\\
HE1122--1648 & 1.88--2.37 & 3507--4098 &35-65\\
HE1347--2457 & 2.05--2.57 & 3711--4352 &50-70\\
HE2217--2818 & 1.88--2.38 & 3503--4107 &35-60\\
J2233--606   & 1.74--2.22 & 3337--3912 &30-50\\
\noalign{\smallskip}
\hline
\end{tabular}
\end{table}

The observed intensity is related to the emitted intensity 
as  $I_{\rm obs} = I_{\rm em} e^{-\tau}$. 
The fluctuations in the observed intensity are thus a superposition 
of the fluctuations of the emitted intensity and those of the
absorption optical depth.  We use here continuum-fitted 
spectra and consider the quantity 
$F= I_{\rm obs}/I_{\rm em}=e^{-\tau}$ (F1 in the notation of Kim et
al. 2003). 
Our  procedure for calculating the flux power spectrum is 
described in detail in K03 and we give here just a short summary.

\begin{figure*}
\center\resizebox{.88\textwidth}{!}{\includegraphics{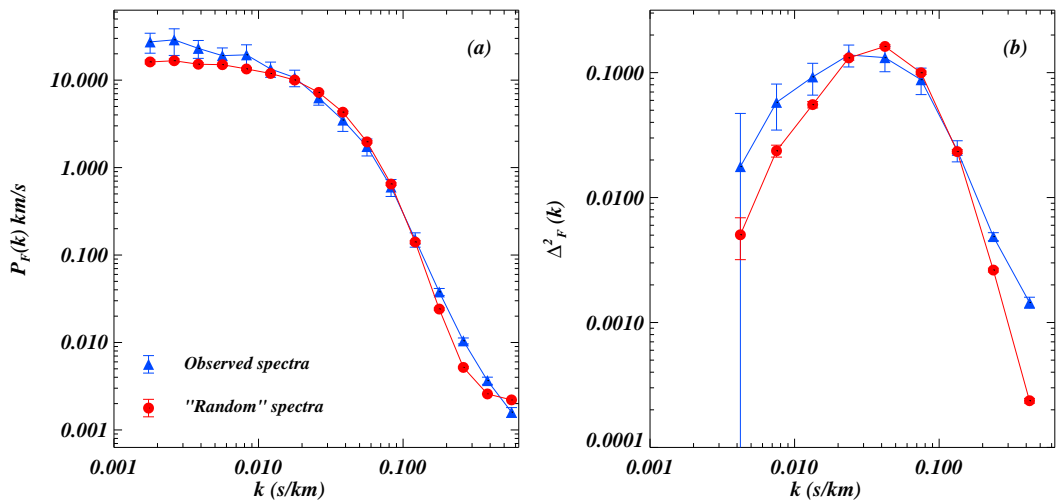}}
\caption{{\protect\footnotesize {{\it Left:} The 1D flux power
spectrum of a sample of observed and randomised spectra. {\it Right:}
``3D'' power spectrum}}}
\label{fig1}
\end{figure*}

The ``3D'' flux 
power spectrum is obtained via numerical differentiation of the 1D 
flux power spectrum, 
\be
P_{\rm F}^{{\rm ``3D''}}(k)= -\frac{2\pi}{k}\frac{d}{dk} P_{{\rm
F}}^{{\rm 1D}}(k).
\ee
The dimensionless  ``3D'' power spectrum is given by,  
\be
\Delta_{\rm F}^2(k)=\frac{1}{2\pi^2}\,k^3\,P_F^{{\rm ``3D''}}(k)\, \;.
\ee 

Jackknife estimates are used to calculate the errors.  
In order to remind the reader that peculiar velocities and thermal
broadening make the flux field anisotropic and that equation  
(2) does not give the true  3D power spectrum we  denote it 
as ``3D''  power spectrum  as in K03.

\subsection {Flux power spectra of observed and randomised absorption 
spectra}

In Fig.~1a we compare the 1D   flux power of the sample of 
observed spectra to that of a sample of randomised spectra. 
For each observed spectrum we have produced  
150 randomised versions. 
At wavenumbers  $0.03$ s/km $<k <0.1$ s/km   
the 1D flux power spectra are remarkably 
similar.
At larger scales the flux power
spectrum becomes close to  constant as expected for a random
distribution of discrete absorption features.
This suggests that the shape of the absorption lines as 
identified by a Voigt profile fitting routine 
dominates  the flux power spectrum over a wide range of scales. 
The discrepancy at  $k>0.1$ s/km is expected as the flux power 
spectrum  of the observed absorption spectra is dominated by 
metal lines at these small scales (K03) which are not 
included in the randomised spectra.  
The  contribution of large scale correlations in the 
density field  is responsible for the  difference at large scales 
but astonishingly these appear to be a relatively small  contribution to the 
overall flux power spectrum. 
To investigate this in more detail  
we plot the corresponding ``3D'' flux power spectra
in Fig 1b.  As suspected 
from inspection of Fig. 1a at the peak   the ``3D'' flux power   
spectra are identical to within the  errors. At larger scales 
the variance of the randomised spectra  is still $\ga$ 50 \% of that
of the observed spectra.  This may explain why 
Cristiani et al. (1995) only found a very weak signal 
for strong absorption systems when they investigated the clustering 
of absorption lines.  It also explains why linear theory 
gives a bad approximation to the ``3D'' flux  power spectrum,  
even at scales $k\sim 0.003-0.001$ s/km (Croft et al. 2002) 
and why numerical simulations are essential for inferring 
the DM power spectrum on these scales.

\section{The effect of strong discrete  absorption lines}

\subsection{Strong absorption lines and  the flux power spectrum}

In Fig.  2a we have splitted the ``3D'' flux power spectrum 
into contributions from lines of different column density ranges. 
Lines of different column density ranges do not add exactly linearly 
but the relative contributions should nevertheless become apparent in this
way. There is a clear trend with larger column density systems contributing
at larger scales. The strongest contribution comes from absorption
lines in the range $13.5<\log(N_{\rm HI}/\cm^{-2})<14.5$. These are lines 
where the ``curve of growth'' which describes the relation between
equivalent width and column density of absorption lines 
changes from the linear to the 
flat regime due to saturation. Such a behaviour can be  understood 
if the different column density ranges contribute with their  total 
equivalent width, $\cal{N}_{\rm lines} \times  {\rm EW}$,  
to the ``3D'' power spectrum. 
The number of lines scales roughly as as ${\cal{N}}_{\rm lines} 
\propto N_{\rm HI}^{-0.5}$.   The contribution to the 
total equivalent width has
therefore a maximum for column densities 
at the transition of the curve of growth
from linear to flat.     
Note that Press et al. (1993) had already demonstrated that 
the rms fluctuation of the flux can be explained with 
randomly distributed lines with a contribution $\propto EW^2$ to 
the 1D variance of the flux.   

\begin{figure*}
\center\resizebox{.88\textwidth}{!}{\includegraphics{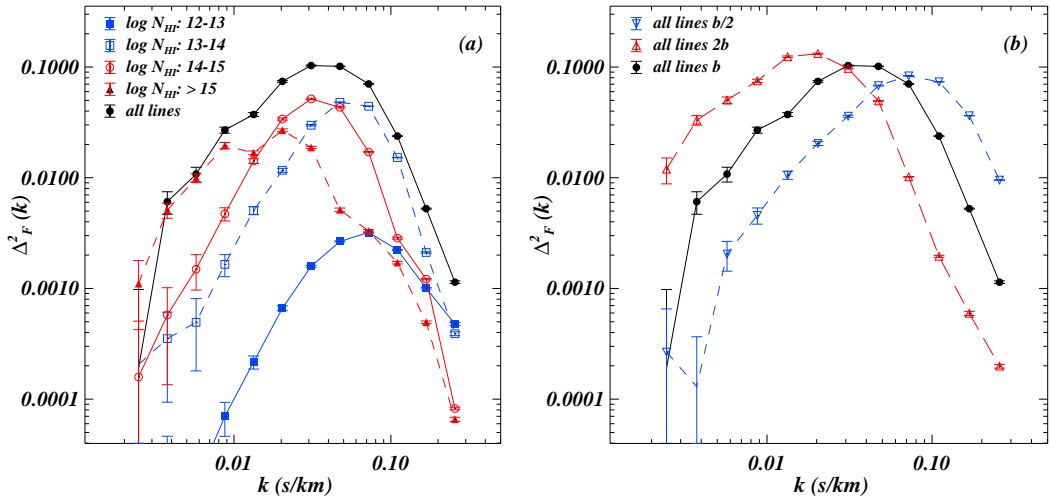}}
\caption{{\protect\footnotesize {{\it Left:} 
Contribution of different column density ranges to the ``3D'' 
flux power spectrum of randomised absorption spectra.  {\it Right:}
The effect of halving and doubling the Doppler parameters.}}}
\label{fig2}
\end{figure*}

To address the question whether it is really the shape of the 
individual Voigt profiles which is responsible for the 
bulk of the flux power spectrum we have also calculated randomised spectra where
we  halved and doubled the Doppler parameter of the lines. 
The resulting ``3D'' power spectra are compared in Figure 2b. Broadening 
and narrowing of the lines leads to an approximately linear shift 
in wavenumber.

\subsection{Strong absorption lines and  the mean flux decrement}

With increasing size of the data sets the mean flux decrement 
of  absorption spectra has been determined reasonably well 
(Press et al. 1993, Rauch et al. 1997, Kim et
al. 2002, Bernardi et al. 2003). 
It is nevertheless  striking that the mean flux decrement 
varies considerably between different QSOs. 
The main 
reason for this is  likely to be the  large contribution of
strong absorption systems. In Fig.~ 3a we have  
used the bottom four of the  spectra listed in Table 1 to   
quantify the contribution 
of absorption lines in different column density ranges to the mean
flux decrement by successively removing low column density lines. 
These four spectra were chosen because they are at similar redshifts
and thus have comparable total mean flux decrement.  
Up to  20 \% of the flux decrement is due to
absorption systems with $\log (N_{\rm HI}/\cm^{-2}) > 15$. 
In Fig.~ 3b we show a similar plot for the {\it rms} fluctuations of 
the flux. About 15-40\% of the flux fluctuations are contributed by 
absorption from strong lines. 
Poisson noise of the rare strong
absorption systems can thus indeed explain the significant 
differences of the mean flux decrement between different
lines-of-sight (Press et al. 1993). 
Unfortunately the sample is too small to test the evolution 
of the contribution of strong absorption lines  with redshift.

\section{Strong absorption systems  and estimates of the 
dark matter power spectrum}

In the last section we had seen that strong absorption 
systems contribute significantly to both flux power spectrum 
and mean flux decrement. This will have profound implications for 
attempts to use numerical simulation together with QSO absorption
spectra to infer amplitude and slope of the DM power spectrum 
with high accuracy. 
Numerical simulations of the \lya forest often 
underpredict the number of strong absorption systems. 
Katz et al.  (1996)  {\it e.g.}  found that in  their 
hydro-simulations 
the number 
of Lyman limit systems falls short by a factor of ten 
compared to the observed number while 
Gardner et al. (2001) find a discrepancy of about a factor five.
For the analysis of \lya flux power spectra numerical simulation 
of the DM distribution which mimic the effect of gas pressure
in an approximate way (so called Hydro-PM simulations)  are widely
used. These simulation have 
rather low resolution and the discrepancy already becomes 
large for lines with  $\log (N_{\rm HI}/\cm^{-2}) > 14$ (Gnedin  1998).      

The lack of strong absorption systems will affect the inferred
matter power spectrum in two ways. 
If large column densities absorption systems contribute significantly  
to the observed flux power spectrum on large scales by their shape their absence
has to be compensated by extra power due to density correlations 
on these scales.  This will require  a shallower slope of the DM 
power spectrum. 
Without detailed numerical simulations it is difficult to estimate how 
large this bias
will be but considering the  large contribution of strong
absorption lines and the  weak dependence of the slope of the
flux power spectrum on the slope of the DM spectrum (Croft et
al. 2002) 
it is
unlikely to be smaller than the current errors of the slope and amplitude 
(McDonald et al. 2002, Croft et al. 2002).   This should thus be a
serious concern for upcoming determinations of the DM power spectrum  
with larger samples of spectra which aim at higher accuracy. 

The second more subtle bias is due to the 
effect of strong absorption systems on the mean flux decrement. 
As  discussed {\it e.g.} by Croft et al. (1998) the amplitude of \lya flux 
power spectrum depends not only the amplitude of the matter power 
spectrum but there is also a strong dependence  on the mean flux 
decrement. 
The amplitude of  the flux power spectrum increases 
with increasing mean flux decrement. 

\begin{figure*}
\center\resizebox{.88\textwidth}{!}{\includegraphics{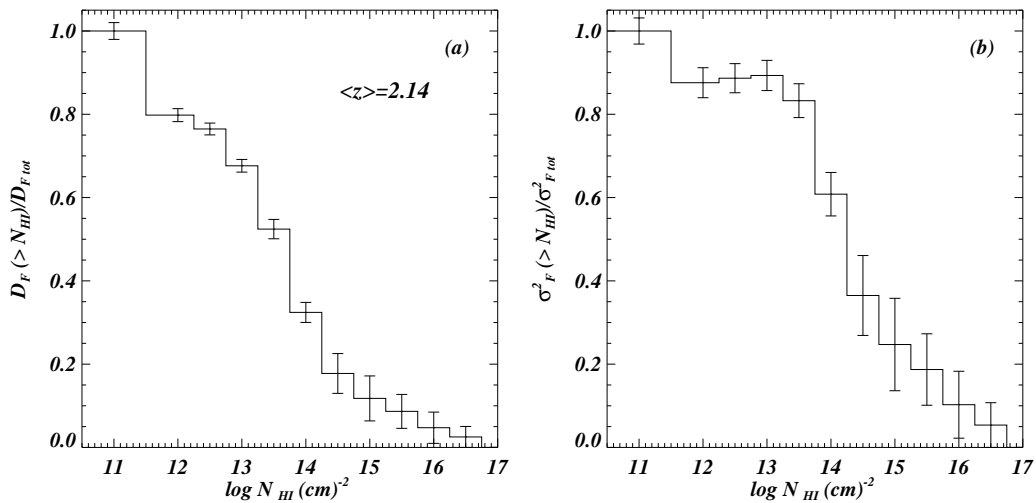}}
\caption{{\protect\footnotesize {{\it Left:} 
Cumulative distribution of the contribution of different column
density ranges to the mean flux decrement of four observed spectra.  
{\it Right:} Same for the {\it rms} amplitude of the flux.}}}
\label{fig3}
\end{figure*}

Analyses  of observed flux power spectra with numerical
simulations therefore normally set the mean flux decrement in the 
artificial spectra produced from numerical simulations to that
of observed spectra. This is a reasonable thing to do if the numerical
simulations are a fair representation of the absorbers responsible for
the flux decrement in observed spectra.  The failure of the
simulation to produce enough strong absorption systems shows,  however,
that this is not the case.  In the case of the hydro-simulations 
this may be related to the fact  that the ionization rate 
required to match the observed mean flux decrement with 
favoured values of the baryon density  and temperature  falls short of
that estimated from QSO surveys and high redshift galaxies 
by a factor 2-4 (Rauch et al. 1997; Haehnelt et al. 2001). 
A population of strong absorption systems may thus
be genuinely missing in numerical simulations.  
In the case of the hydro-PM simulations 
the insufficient spatial resolution is the obvious reason 
for the much  more severe lack of strong absorption systems.

Calibration of numerical simulations 
which underreproduce observed strong absorption systems 
using the observed mean flux decrement is thus clearly not the right 
thing to do. If it is nevertheless done the amplitude 
of the matter  power spectrum inferred from the \lya flux power
spectrum will be systematically biased  low. The simulated
spectra reproduce then  the observed  flux fluctuations with 
a larger flux decrement and a smaller matter fluctuation amplitude.     
This effect is not small. Seljak, McDonald \& Makarov (2003)  found
that a 20 \% reduction of the effective optical depth used to
calibrate the simulated spectra leads to a factor 2 increase of  the
inferred DM fluctuation amplitude.  
Note that if the inferred DM fluctuation amplitude at
small scales is biased low a combined analysis with other data on
larger scales  (CMB, galaxy surveys)
will  lead to an underestimate of the   spectral index. Alternatively,  
a spurious running of the spectral index may be inferred.

\section{Conclusions}

We used a  sample of high-resolution,  high signal-to-noise QSO 
absorption spectra taken with the ESO-UVES spectrograph 
for which  line lists of hydrogen and metal absorption systems 
are available to investigate the effect of discrete absorption systems 
on  \lya flux power spectra. 
The basic shape and amplitude  of the \lya flux power spectrum 
is well reproduced by a random superposition of Voigt profiles.  
Only at wavelengths larger than  $k < 0.03 $ s/km  a weak clustering 
signal appears to be detected.  However, even there  the  contribution from 
strong absorption systems is still $\ga 50 \%$. 
The contribution of strong
absorption systems to the mean flux decrement is also large,  
up to $20 \%$ for absorption systems with 
$\log (N_{\rm HI}/\cm^{-2}) > 15 $.  Simulated absorption 
spectra often  underreproduce the number of strong absorption
systems.  The use of such   
simulations calibrated to the observed mean flux 
for an analysis of \lya forest flux power spectrum 
is likely to lead to an underestimate of the  amplitude and  initial
slope of the inferred DM power spectrum.

\section*{Acknowledgments.} 
This work was supported by the European Community Research and
Training Network ``The Physics of the Intergalactic Medium''.
We would like to thank ESO for  making publicly available  a superb set of QSO
absorption spectra. MV thanks Cristiano Porciani and Sabino Matarrese
for useful discussions.


\begin{thebibliography}{}
\bibitem[]{} Bernardi M. et al., 2003, AJ, 115, 32 
\bibitem[]{} Cristiani S.,  D'Odorico S.,  Fontana A., Giallongo E.,  
 Savaglio S., 1995, MNRAS, 273, 1016
\bibitem[]{} Croft R. A. C., Weinberg D. H., Katz N., Hernquist L.,
1998, ApJ, 495, 44
\bibitem[]{} Croft R.A.C., Weinberg D. H., Pettini M., Hernquist L.,
Katz N.,  1999, ApJ, 520, 1
\bibitem[]{} Croft R.A.C. Croft, Weinberg D. H., Bolte M., Burles S.,
Hernquist L., Katz N.,Kirkman D., Tytler D., 2002, ApJ, 581, 20
\bibitem[]{} Gardner J.P. , Katz N., Hernquist L., Weinberg D.H.,
2001, ApJ, 559, 131
\bibitem[]{} Gnedin N.Y., 1998, MNRAS, 299, 392 
\bibitem[]{} Haehnelt M.G., Madau P., Kudritzki R.P., Haardt F., 
2001, ApJ, 549, L151 
\bibitem[]{} Katz N., Weinberg D.H., Hernquist L., 1996, ApJS, 105, 19
\bibitem[]{} Kim, T.-S., Cristiani, S., D'Dodorico, S. 2001, A\&A, 373, 757
\bibitem[]{} Kim T.-S., Carswell, R. F., Cristiani, S., 
D'Odorico, S., Giallongo E., 2002, MNRAS, 335, 555
\bibitem[]{} Kim T.-S., Viel M., Haehnelt M.G., Cristiani S., Carswell
B., 2003, submitted to MNRAS (K03)
\bibitem[]{} Mandelbaum, R.,  McDonald, P., Seljak, U.,   Cen, R. ,
2003, astro-ph/0302112 
\bibitem[]{} McDonald P., Miralda-Escud\'e J., Rauch M., Sargent W.L.,
Barlow T.A., Cen R., Ostriker J.P., 2000, ApJ, 543, 1
\bibitem[]{} Press W.H., Rybicki G.B., Schneider D.P., 1993, ApJ, 414, 64
\bibitem[]{} Rauch M., 1998, ARA\&A, 36, 267
\bibitem[]{} Rauch M., Miralda-Escude, J.,  Sargent W.L.W., Barlow T.A.,
 Weinberg D.H., Hernquist L., Katz N.,  Cen R.,  Ostriker J.P., 1997, ApJ,
489, 7
\bibitem[]{} Seljak U., McDonald P., Makarov A., 2003, astro-ph/0302571
\bibitem[]{} Spergel D.N. et al. 2003, astro-ph/0302209
\bibitem[]{} Verde L. et al. 2003, astro-ph/0302218 
\bibitem[]{} Zuo L., Bond J.R., 1994, ApJ, 423, 73  
\end{thebibliography}
\end{document}